\documentclass[12pt,a4paper]{article}

\usepackage{amsmath}
\usepackage{amsfonts}

\numberwithin{equation}{section}

\newcommand{\nc}{\newcommand}
\nc{\rnc}{\renewcommand}

\rnc{\title}[1]{\large\mbox{}\\ \mbox{}\\ \mbox{}\\
     \textbf{#1}\bigskip\medskip\\}
\rnc{\author}[1]{\large #1\\ \smallskip}
\nc{\address}[1]{{\narrower\normalsize\it #1\\}\bigskip}

\nc{\be}{\begin{equation}}
\nc{\round}[1]{\left(#1\right)}
\nc{\squareb}[1]{\left[#1\right]}
\nc{\curly}[1]{\left\{#1\right\}}
\nc{\abs}[1]{\left|#1\right|}
\nc{\pointy}[1]{\left\langle #1 \right\rangle}
\nc{\lsum}[3]{\sum_{#1=#2}^{#3}}

\nc{\Wt}[4]{W
\begin{pmatrix}
#4 & #3\\
#1 & #2
\end{pmatrix}}

\nc{\Kt}[3]{K
\begin{pmatrix}
#2 &
\begin{matrix}
#3\\
#1
\end{matrix}
\end{pmatrix}}

\nc{\Db}{D_{\text{b}}}
\nc{\Ds}{D_{\text{s}}}
\nc{\Df}{D_{\text{f}}}
\nc{\Tb}{T_{\text{b}}}
\nc{\ks}{\kappa_{\text{s}}}
\nc{\kb}{\kappa_{\text{b}}}
\nc{\fsing}{f_{\text{sing}}}
\nc{\fsinglr}{\fsing^{\text{L},\text{R}}}
\nc{\Cb}{C_{\text{b}}}
\nc{\Cs}{C_{\text{s}}}
\nc{\als}{\alpha_{\text{s}}}
\nc{\fb}{f_{\text{b}}}
\nc{\fs}{f_{\text{s}}}

\nc{\xil}{\xi_{\text{L}}}
\nc{\xir}{\xi_{\text{R}}}
\nc{\al}{a_{\text{L}}}
\nc{\ar}{a_{\text{R}}}
\nc{\kslr}{\ks^{\text{L},\text{R}}}
\nc{\epsl}{\eps^{\text{L}}}
\nc{\epsr}{\eps^{\text{R}}}
\nc{\epslr}{\eps^{\text{L},\text{R}}}

\nc{\thf}{\vartheta_1}
\nc{\thfr}{\vartheta_4}

\nc{\lam}{\lambda}
\nc{\Lam}{\Lambda}
\nc{\Om}{\Omega}
\nc{\gam}{\gamma}
\nc{\eps}{\epsilon}
\nc{\veps}{\varepsilon}
\nc{\kap}{\kappa}
\nc{\m}{\mu}

\nc{\Dm}{\mathbf{D}}
\nc{\av}{\mathbf{a}}
\nc{\bv}{\mathbf{b}}

\DeclareMathOperator{\Real}{Re}

\nc{\sm}[1]{{\scriptstyle #1}}
\nc{\ssm}[1]{{\scriptscriptstyle #1}}
\nc{\spos}[2]{\makebox(0,0)[#1]{$\sm{#2}$}}
\nc{\dl}[3]{\put(#1,#2){\makebox(#3,0){\dotfill}}}
\rnc{\d}[2]{\put(#1,#2){\spos{}{\bullet}}}
\nc{\dd}[3]{\multiput(#1,#2)(0,1){#3}{\spos{}{\bullet}}}

\begin{document}

\begin{center}
\title{Surface Free Energies and Surface Critical Behaviour\\
of the ABF Models with Fixed Boundaries\footnote{To be published in the
proceedings of \emph{Yang-Baxter Equations and Related Topics}, Tianjin, China,
August 1995.}}
\author{
David L. O'Brien\footnote{E-mail address: \texttt{dlo@maths.mu.oz.au}},
Paul A. Pearce\footnote{E-mail address: \texttt{pap@maths.mu.oz.au}.  On leave
from The University of Melbourne.} and
Roger E. Behrend\footnote{E-mail address: \texttt{reb@maths.mu.oz.au}}}
\address{\footnotemark[2]${}^,$\footnotemark[4]Department of Mathematics, The
University of Melbourne,\\ Parkville, Victoria 3052, Australia\\
and\\
\footnotemark[3]Physikalisches Institut der Universit\"{a}t Bonn,\\
Nu\ss allee 12, D-53115 Bonn, Germany.}

\begin{abstract}
\noindent
In a previous paper, we introduced reflection equations for
interaction-round-a-face (IRF) models and used these to construct commuting
double-row transfer matrices for solvable lattice spin models with fixed
boundary conditions. In particular, for the Andrews-Baxter-Forrester (ABF)
models, we derived special functional equations satisfied by the eigenvalues of
the commuting double-row transfer matrices. Here we introduce a generalized
inversion relation method to solve these functional equations for the surface
free energies. Although the surface free energies depend on the boundary spins
we find that the associated surface critical exponent $\als=(7-L)/4$ is
independent of the choice of boundary.
\end{abstract}
\end{center}

\section{Introduction}

We have recently demonstrated how, by generalizing the work of Sklyanin
\cite{Sk88}, fixed boundary conditions may be imposed upon
interaction-round-a-face (IRF) models whilst preserving solvability
\cite{BPO95}.  Specifically, we have shown that the Yang-Baxter equations and
the boundary reflection equations imply commutativity of double-row transfer
matrices.  Related results have also been obtained by the authors of
\cite{Kul95}
and~\cite{AW95}.

Furthermore, we have found solutions to the reflection equations for
the Andrews-Baxter-Forrester (ABF) models \cite{ABF84}.  Consideration of the
fusion
\cite{DJMO86,DJMO87,DJKMO87,DJKMO88}
of this model with fixed boundaries leads to functional relations for
the double-row transfer matrices, and hence their eigenvalues.  In this paper
we
solve these functional equations in the thermodynamic limit to obtain the
surface free energy, from which we determine the surface critical exponent
$\als$.

In the remainder of this section we summarize the results of \cite{BPO95}
which are needed for our calculation.  For further details we refer the reader
to
this reference.  In Section~2 we demonstrate the
solution of the inversion relation for the bulk free energy, and then
generalize this method to the surface free energy.

\subsection{The ABF models with fixed boundaries}

The ABF models \cite{ABF84} are restricted solid-on-solid models in which sites
on the lattice take values in the set $\{1,2,3,\dots,L\}$ subject to the
condition that the values of sites adjacent on the lattice must differ
by $\pm1$.  The Boltzmann weights depend on a \emph{crossing parameter}
$\lam=\pi/(L+1)$, and a \emph{spectral parameter} $u$.  In the regimes of
interest we have $0<u<\lam$. The non-zero face weights are given by
\begin{align}
\Wt{a}{a\mp1}{a}{a\pm1}&=\frac{\thf(\lam-u)}{\thf(\lam)}\\
\Wt{a\mp1}{a}{a\pm1}{a}&=
\round{\frac{\thf((a-1)\lam)\thf((a+1)\lam)}{\thf^2(a\lam)}}^{1/2}\,
\frac{\thf(u)}{\thf(\lam)}\\
\Wt{a\pm1}{a}{a\pm1}{a}&=\frac{\thf(a\lam\pm u)}{\thf(a\lam)}.
\end{align}
The $\thf(u)=\thf(u,p)$ are elliptic theta functions
with nome $p$.  We define
$p=\exp(-\veps)$ with $\veps>0$, corresponding to regime III\@.
The critical limit is
$p\to 0$. The boundary weights depend on an additional arbitrary complex
parameter $\xi$, which may be different for the left and right boundaries.  The
non-zero boundary weights are
\begin{equation}
\Kt{a}{a\pm1}{a}=\round{\frac{\thf((a\pm1)\lam)}{\thf(a\lam)}}^{1/2}\,
\frac{\thf(u\pm\xi)\thf(u\mp a\lam\mp\xi)}{\thf^2(\lam)}.
\label{boundwts}
\end{equation}
There is another form of the boundary weights which, at criticality, is
independent of $u$ and $\xi$
\begin{equation}
\Kt{a}{a\pm1}{a}=\round{\frac{\thf((a\pm1)\lam)}{\thf(a\lam)}}^{1/2}\,
\frac{\thfr(u\pm\xi)\thfr(u\mp a\lam\mp\xi)}{\thfr^2(\lam)}.\label{tlbw}
\end{equation}
This expression is obtained from \eqref{boundwts} simply by making a complex
shift in $\xi$.  However, in this paper we will take $\xi$ to be real and
consider only the two forms of the boundary weights \eqref{boundwts} and
\eqref{tlbw}. From the face weights and boundary weights we construct a
double-row transfer matrix $\Dm(u)$.  For a lattice of width $N$, the entry of
the transfer matrix corresponding to the rows of spins
$\av=\{a_1,\dots,a_{N+1}\}$ and
$\bv=\{b_1,\dots,b_{N+1}\}$ is defined diagrammatically by
\setlength{\unitlength}{12mm}
\begin{equation*}
\pointy{\av|\Dm(u)|\bv}
=\raisebox{-1.4\unitlength}[1.6\unitlength][
1.4\unitlength]{\begin{picture}(6.2,3)(0.4,0)
\multiput(0.5,0.5)(6,0){2}{\line(0,1){2}}
\multiput(1,0.5)(1,0){3}{\line(0,1){2}}
\multiput(5,0.5)(1,0){2}{\line(0,1){2}}
\multiput(1,0.5)(0,1){3}{\line(1,0){5}}
\put(1,1.5){\line(-1,2){0.5}}\put(1,1.5){\line(-1,-2){0.5}}
\put(6,1.5){\line(1,2){0.5}}\put(6,1.5){\line(1,-2){0.5}}
\put(0.5,0.45){\spos{t}{a_1}}\put(1,0.45){\spos{t}{a_1}}
\put(2,0.45){\spos{t}{a_2}}\put(3,0.45){\spos{t}{a_3}}
\put(5,0.45){\spos{t}{a_N}}\put(6,0.45){\spos{t}{a_{N\!+\!1}}}
\put(6.7,0.45){\spos{t}{a_{N\!+\!1}}}
\put(0.5,2.6){\spos{b}{b_1}}\put(1,2.6){\spos{b}{b_1}}
\put(2,2.6){\spos{b}{b_2}}\put(3,2.6){\spos{b}{b_3}}
\put(5,2.6){\spos{b}{b_N}}\put(6,2.6){\spos{b}{b_{N\!+\!1}}}
\put(6.7,2.6){\spos{b}{b_{N\!+\!1}}}
\put(1.05,1.45){\spos{tl}{c_1}}\put(2.05,1.45){\spos{tl}{c_2}}
\put(3.05,1.45){\spos{tl}{c_3}}\put(4.99,1.45){\spos{tr}{c_N}}
\put(5.99,1.45){\spos{tr}{c_{N\!+\!1}}}
\multiput(1.5,1)(1,0){2}{\spos{}{u}}\put(5.5,1){\spos{}{u}}
\multiput(1.5,2)(1,0){2}{\spos{}{\lam\!-\!u}}
\put(5.5,2){\spos{}{\lam\!-\!u}}
\put(0.71,1.5){\spos{}{\lam\!-\!u}}\put(6.29,1.5){\spos{}{u}}
\multiput(0.5,0.5)(0,2){2}{\makebox(0.5,0){\dotfill}}
\multiput(6,0.5)(0,2){2}{\makebox(0.5,0){\dotfill}}
\multiput(1,1.5)(1,0){3}{\spos{}{\bullet}}
\multiput(5,1.5)(1,0){2}{\spos{}{\bullet}}
\end{picture}}
\end{equation*}
The solid spins $\{c_1,\dots,c_{N+1}\}$ are summed over. As the boundary
weights
are diagonal, we must have $a_1=b_1$ and
$a_{N+1}=b_{N+1}$.  Furthermore, these boundary spins, which we will call $\al$
and $\ar$, are fixed to the same values for all entries in the transfer matrix.
The parameters $\xil$ and $\xir$ are similarly fixed for all entries.  Defined
in
this way, the double-row transfer matrix exhibits the crossing symmetry
\begin{equation}
\Dm(\lam-u)=\Dm(u).
\label{Dcross}
\end{equation}
More importantly, however, the double-row transfer matrices form a commuting
family,
\begin{equation}
\Dm(u)\Dm(v)=\Dm(v)\Dm(u).
\end{equation}
This implies that the eigenvectors of $\Dm(u)$ are independent of $u$, so that
functional equations satisfied by the transfer matrix are also satisfied by its
eigenvalues.  In particular, all eigenvalues of the transfer matrix satisfy the
crossing symmetry \eqref{Dcross}. It should be emphasized that all the matrices
in a commuting family share the same boundary spins $\al$ and $\ar$, and the
same
values of $\xil$ and $\xir$.

The values $\xi=\pm\lam/2$ deserve special mention, since for these choices the
isotropic lattice, $u=\lam/2$, has all boundary spins fixed.  This is easily
seen
from the definition \eqref{boundwts}, as, for fixed $a$, only one of the
choices
$a\pm1$ gives a non-zero boundary weight.  The non-zero boundary weights then
contribute only a constant factor to each entry the transfer matrix.  Aside
from this trivial factor, the lattice exhibits pure fixed boundary conditions,
with boundary spins alternating $\{a,a+1,a,a+1,\dots\}$ or
$\{a,a-1,a,a-1,\dots\}$.

Just as in the case of periodic boundary conditions, the face weights and
boundary weights may be fused \cite{DJMO86,DJMO87,DJKMO87,DJKMO88} to form new
solvable models with fixed boundary conditions.  The functional equations which
result have the same form as in the periodic case, with the addition of
some order~$1$ factors related to the boundary.

\section{Functional equations}

It has been shown in \cite{BPO95} that the eigenvalues of the ABF models
with fixed boundary conditions at fusion level $1\times q$ satisfy the
inversion identity hierarchy
\begin{equation}
s_{q-1}s_{q+1}D^q(u)D^q(u+\lam)=\gamma^q_{q-1}
s_{-1}s_{2q+1}f_{-1}f_q+
s^2_qD^{q-1}(u+\lam)D^{q+1}(u),
\end{equation}
where $1\le q\le L-1$, and
\begin{equation}
s_k=\thf(2u+(k-1)\lam), \qquad
\gamma^r_k=\alpha^r_k\beta^r_k,\qquad
f_k=(-1)^N\squareb{\frac{\thf(u+k\lam)}{\thf(\lam)}}^{2N}.
\end{equation}
In terms of the function
\begin{equation}
\theta^r_k(u)=\prod_{j=0}^{r-1}\frac{\thf(u+(k-j)\lam)}{\thf(\lam)},
\label{thfdef}
\end{equation}
we can write $\alpha^r_k$ and $\beta^r_k$ as
\begin{equation}
\begin{split}
\alpha^r_k(u)&=\theta^r_k(u-\xil)\theta^r_k(u+\xil)
\theta^r_k(u-\xir)\theta^r_k(u+\xir) \\
\beta^r_k(u)&=\theta^r_{k-\al}(u-\xil)\theta^r_{k+\al}(u+\xil)
\theta^r_{k-\ar}(u-\xir)\theta^r_{k+\ar}(u+\xir).
\end{split}
\end{equation}
If the second form of the boundary weights \eqref{tlbw} is used, the $\thf$
functions in \eqref{thfdef} should be changed to $\thfr$ functions.
The closure of the inversion identity hierarchy is governed by the
conditions
\begin{equation}
D^{-1}(u)=D^L(u)=0,\quad D^0(u)=f_{-1},\quad
D^{L-1}(u)=(-1)^Nf_{-2}\alpha^L_{L-2}/\beta^1_{-2}.
\end{equation}
If we write the eigenvalues $D^q(u)$ with their bulk and surface terms
separated
\begin{equation}
D^q(u)\sim\Db^q(u)\Ds^q(u)\quad\text{as $N\to\infty$,}
\label{sep}
\end{equation}
then, by virtue of the inversion relation for the fused face weights
\cite{DJMO86},
$\Db^q(u)$ satisfies the functional relation
\begin{equation}
\Db^q(u)\Db^q(u+\lam)=f_{-1}f_q,
\label{bulkinv}
\end{equation}

\subsection{The bulk free energy}

Equation \eqref{bulkinv} implies that, between the inversion points
$u=(1-q)\lam/2$ and \mbox{$(3-q)\lam/2$},
the bulk partition function per site
satisfies the functional equation
\begin{equation}
\kb^q(u)\kb^q(u+\lam)=\frac{\thf(\lam-u)\thf(u+q\lam)}{\thf(\lam)^2}.
\label{kapinv}
\end{equation}
This is the same inversion relation as
in the case of periodic conditions, which is to be expected since the
boundary conditions should not affect the bulk behaviour.  Equation
\eqref{kapinv} has been solved previously \cite{DJKMO87,DJKMO88,BR89}, but we
include the solution here for completeness, and as an introduction to the
generalized methods that follow.  We use the standard techniques developed by
Baxter
\cite{Bax82}. The
assumption that the solutions are analytic between the inversion points, and
that
they may be analytically continued a small distance outside the strip, along
with the relation
\eqref{kapinv} and the crossing symmetry
\begin{equation}
D^q((2-q)\lam-u)=D^q(u)
\label{fusedcross}
\end{equation}
uniquely
determines the free energies.  The assumption of analyticity may be justified
by studying the zeros of the largest eigenvalue $D^q(u)$ for large finite $N$.
In the critical case, it is seen that for the fusion level $1\times q$, the
strip $-q\lam/2<\Real(u)<(4-q)\lam/2$ is free of order~$N$ zeros \cite{KP92}.
Furthermore, there can be no poles inside this strip, since for finite $N$ the
independence of the eigenvectors on $u$ implies that the eigenvalues are simply
linear combinations of products of Boltzmann weights.  Since none of the
Boltzmann weights have poles, neither can the eigenvalues.
This assumption of analyticity implies that the logarithms of the partition
functions may be expanded in a Laurent series in powers of
$\exp(2\pi u/\veps)$,
\begin{equation}
\ln\kb^q(u)=\sum_{k=-\infty}^{\infty}c_ke^{2k\pi u/\veps}.
\end{equation}
We rewrite the right hand side of \eqref{kapinv} using the ``conjugate
modulus'' transformation, which, in terms of the function
\begin{equation}
E(x,p)=\prod_{n=1}^\infty(1-p^{n-1}x)(1-p^nx^{-1})(1-p^n),
\end{equation}
is given by
\begin{align}
\thf(u,p)&=\round{\frac{\pi}{\veps}}^{1/2}e^{-(u-\pi/2)^2/\veps}\,
E(e^{-2\pi u/\veps},\tilde{q}^2)\\
\thfr(u,p)&=\round{\frac{\pi}{\veps}}^{1/2}e^{-(u-\pi/2)^2/\veps}\,
E(-e^{-2\pi u/\veps},\tilde{q}^2)
\end{align}
where $p=\exp(-\veps)$ and $\tilde{q}=\exp(-\pi^2/\veps)$ are
conjugate nomes.  With both sides of \eqref{kapinv} expanded in
powers of $\exp(2\pi u/\veps)$, we match coefficients and impose the crossing
symmetry \eqref{fusedcross} to obtain the solution
\begin{multline}
\ln\kb^q(u)=c_0(u)+
\sum_{k=1}^{\infty}
\frac{\cosh[(\pi-2\lam)\pi k/\veps]}{k\sinh(\pi^2 k/\veps)} \\
-\sum_{k=1}^{\infty}
\frac{\cosh[(\pi-(q+1)\lam)\pi k/\veps]\cosh[((2-q)\lam-2u)\pi k/\veps]}
{k\sinh(\pi^2 k/\veps)\cosh(\lam\pi k/\veps)},
\label{bulkfe}
\end{multline}
where
\begin{equation}
c_0(u)=\frac{1}{2\veps}\squareb{(\pi-q\lam)(q-1)\lam+2u((2-q)\lam-u)}.
\end{equation}
The analytic continuation of this function to the region
$-q\lam/2<\Real(u)<(4-q)\lam/2$ gives the bulk behaviour of the largest
eigenvalue of the transfer matrix
\begin{equation}
\Db^q(u)\sim[\kb^q(u)]^{2N}\quad\text{as $N\to\infty$.}
\label{bf}
\end{equation}
{}From \eqref{bulkfe} it is easy to show that inside
the interval $-q\lam/2<u<(2-q)\lam/2$,
\begin{equation}
\kb^{q-1}(u+\lam)\kb^{q+1}(u)=\kb^q(u)\kb^q(u+\lam)\quad\text{when $q>1$.}
\label{bf1}
\end{equation}
When $q=1$, it can also be shown that inside $-\lam/2<u<\lam/2$,
\begin{equation}
\ln\abs{\frac{\kb^{q-1}(u+\lam)\kb^{q+1}(u)}
{\kb^q(u)\kb^q(u+\lam)}}=-\frac{\pi}{2\veps}(\lam-2|u|)
-\sum_{k=1}^\infty\frac{\sinh(\lam-2|u|)\pi k/\veps}
{k\cosh(\lam \pi k/\veps)},
\end{equation}
so that
\begin{equation}
\abs{\kb^{q-1}(u+\lam)\kb^{q+1}(u)}<\abs{\kb^q(u)\kb^q(u+\lam)}\quad\text{when
$q=1$.}
\label{bf2}
\end{equation}
Putting together equations \eqref{kapinv}, \eqref{bf}, \eqref{bf1} and
\eqref{bf2}, we therefore obtain
\begin{equation}
\lim_{N\to\infty}\round{\frac{\Db^{q-1}(u+\lam)\Db^{q+1}(u)}
{f_{-1}(u)f_q(u)}}=
\begin{cases}
0 & \text{when $q=1$,}\\
1 & \text{when $q>1$,}
\end{cases}
\label{bulkbehav}
\end{equation}
which is consistent with the critical bulk behaviour described in \cite{KP92}.
We need this result for the derivation that follows.

\subsection{The surface free energy}

Recalling the separation of the bulk and surface terms \eqref{sep}, and
using the functional equation \eqref{bulkinv} and the bulk behaviour of the
eigenvalues \eqref{bulkbehav}, the inversion identity hierarchy in the
thermodynamic limit becomes, for
$-q\lam/2<\Real(u)<(2-q)\lam/2$,
\begin{equation}
s_{q-1}s_{q+1}\Ds^q(u)\Ds^q(u+\lam)=
\begin{cases}
\gamma^1_0s_{-1}s_3 & \text{if $q=1$,}\\
\gamma^q_{q-1}s_{-1}s_{2q+1}+
s^2_q\Ds^{q-1}(u+\lam)\Ds^{q+1}(u) & \text{otherwise.}
\end{cases}
\label{bfushier}
\end{equation}
In the case of the unfused model, $q=1$, we therefore have an inversion
relation
which allows us to calculate the surface free energy\footnote{In the general
case of $p\times q$ fusion, an analogous functional equation is derived
when $q=p$.}.  Explicitly,
\begin{equation}
\ks(u)\ks(u+\lam)=\frac{\thf(2\lam-2u)\thf(2\lam+2u)}
{\thf(\lam-2u)\thf(\lam+2u)}\,\epsl(u)\epsr(u), \label{boundinv}
\end{equation}
where $\epsl(u)$ and $\epsr(u)$ are defined by, for $a$ and
$\xi$ corresponding to the left or right boundary as appropriate,
\begin{equation}
\epslr(u)=\theta^1_0(u-\xi)\theta^1_0(u+\xi)
\theta^1_{-a}(u-\xi)\theta^1_a(u+\xi).
\end{equation}
In the unfused case, the crossing symmetry of the eigenvalues implies that
\begin{equation}
\ks(\lam-u)=\ks(u).
\end{equation}
The solution of the inversion relation for the surface free energy
proceeds in a similar fashion as that for the bulk free energy, although
we must justify the assumption of analyticity separately.  There can of
course be no poles between the inversion points for the same reason as there
are none in the bulk.  However, we are now concerned about order~$1$ zeros
inside $0<\Real(u)<\lam$.  Certainly zeros occur on the line $\Real(u)=\lam/2$,
but, guided by the derivation of conformal weights for periodic boundary
conditions \cite{KP92}, we associate these zeros with finite-size (order~$1/N$)
corrections rather than surface effects.

Numerical studies show that zeros do occur on the real $u$ axis
inside the strip
$-\lam/2<\Real(u)<3\lam/2$, but never inside the strip $0<\Real(u)<\lam$.  For
certain values of $\xi$, zeros occur at the inversion points $u=0$ and $\lam$.
These values are determined by the zeros of $\epslr(u)$, and are found to be
$\xi=0$,
$-a\lam$ and $(L+1-a)\lam$.  For all other values of $\xi$ (up to periodicity),
the interval between the inversion points is free of zeros.  We therefore
conclude that the surface partition function per site, $\ks(u)$, is analytic in
this region.

With this
assumption of analyticity, and the imposition of the crossing symmetry, the
solution of \eqref{boundinv} is
\begin{equation}
\begin{split}
\ln\ks(u)&=2\lsum{k}{1}{\infty}\frac{\sinh[(\pi-3\lam)\pi k/\veps]
\sinh(\lam\pi k/\veps)\cosh[2(\lam-2u)\pi k/\veps]}
{k\sinh(\pi^2k/\veps)\cosh(2\lam\pi k/\veps)} \\
&\quad+(\pi-3\lam)\lam/\veps+\ln\ks^{\text{L}}(u)+\ln\ks^{\text{R}}(u),
\end{split}
\label{bfe}
\end{equation}
where $\ln\ks^{\text{L}}$ and $\ln\ks^{\text{R}}$ are given for generic $a$ and
$\xi$ by
\begin{multline}
\ln\kslr(u)=c_0(u)+
2\lsum{k}{1}{\infty}\frac{\cosh[(\pi-2\lam)\pi k/\veps]}
{k\sinh(\pi^2k/\veps)}\\
\quad-2\lsum{k}{1}{\infty}\frac{\cosh[(a\lam+\xi-|\xi|)\pi k/\veps]
\cosh[(\pi-a\lam-\xi-|\xi|)\pi k/\veps]
\cosh[(\lam-2u)\pi k/\veps]}
{k\sinh(\pi^2k/\veps)\cosh(\lam\pi k/\veps)}
\label{xipart}
\end{multline}
and
\begin{equation}
c_0(u)=\frac{1}{\veps}\squareb{(a\lam+\xi)(\pi-2\xi)+(|\xi|-2\lam)\pi+
(2-a^2)\lam^2+2u(\lam-u)}.
\end{equation}
In deriving this expression, $\xi$ is assumed to satisfy the inequality
\be
-\pi<(1-a)\lam<\xi<(L-a)\lam<\pi.
\end{equation}
We note that, as one would expect, the height reversal transformation $a\to
L+1-a$ and
$\xi\to-\xi$ leaves \eqref{xipart} unchanged.
If the second form of the boundary weights \eqref{tlbw} is used, each term in
the sums of \eqref{xipart} should be multiplied by $(-1)^k$, which alters the
critical behaviour.

The temperature variable $t$ is identified in \cite{ABF84} to be given by
$t=p^2$.  The behaviour in the critical limit $t\to0^+$ is
found by applying the Poisson summation formula to the above expression for
$\ln\ks$ \eqref{bfe}.  We find that the leading-order singularities of $\ln\ks$
have the form
\begin{equation}
\fsing\sim
\begin{cases}
t^{\pi/4\lam}&\text{if $L\equiv0$ or $1\pmod{4}$}, \\
t^{\pi/4\lam}\ln t&\text{if $L\equiv3\pmod{4}$}.
\end{cases}
\end{equation}
When $L\equiv2\pmod{4}$, $\ln\ks$ is regular.  The leading singularity
of $\ln\kslr$ is in general of higher order than that of $\ln\ks$.  In
addition, $\ln\kslr$ is regular in the following situations:
\begin{equation*}
\text{$L$ even}
\begin{cases}
\xi=k\lam,\quad k\in\mathbb{Z}\\
\text{$\xi>0$ and $a$ odd}\\
\text{$\xi<0$ and $a$ even}
\end{cases}
\qquad
\text{$L$ odd}
\begin{cases}
\xi=(2k+1)\lam/2,\quad k\in\mathbb{Z}\\
\text{$a$ odd}
\end{cases}
\end{equation*}
These exceptions aside, $\ln\kslr$ has the leading-order singularities
\begin{equation}
\fsinglr\sim
\begin{cases}
t^{\pi/2\lam}&\text{if $L$ is even,} \\
t^{\pi/2\lam}\ln t&\text{if $L$ is odd.}
\end{cases}
\end{equation}
When the second form of the boundary weights \eqref{tlbw} is used, the
leading-order singularities are, aside once again from the above exceptions,
\begin{equation}
\fsinglr\sim
\begin{cases}
t^{\pi/4\lam}&\text{if $L$ is even,} \\
t^{\pi/4\lam}\ln t&\text{if $L$ is odd.}
\end{cases}
\end{equation}
In this case the function $\ln\kslr$ vanishes at criticality.

The surface critical exponent $\als$ is defined in analogy with the bulk
critical exponent $\alpha$ \cite{Binder,Diehl},
\begin{equation}
\Cb \sim \abs{t}^{-\alpha}\qquad
\Cs \sim \abs{t}^{-\als},
\end{equation}
with the specific heats
\begin{equation}
\Cb=\frac{\partial^2 \fb}{\partial t^2}\qquad
\Cs=\frac{\partial^2 \fs}{\partial t^2}.
\end{equation}
Our result for the surface critical exponent $\als$ is therefore
\begin{equation}
\als=\frac{7-L}{4}.
\end{equation}
The ABF model with $L=3$ corresponds to the two-dimensional Ising model.  In
this case we have
\begin{equation}
\fsing\sim t\ln t,
\end{equation}
in agreement with the calculations of McCoy and Wu \cite{McCW73} for free
boundary conditions, but with a lattice rotated by $45^\circ$ with respect to
the one considered by them.

\section{Conclusion}

{}From the inversion identity hierarchy, and from the known solution of the
bulk
free energy, we have derived an inversion relation for the surface free energy
of the ABF models with fixed boundary conditions.  We have solved this
inversion relation, subject to justifiable analyticity assumptions, and thus
obtained an expression for the surface free energy.  Finally, we have analysed
the critical behaviour of the surface free energy to obtain the surface
critical exponent $\als$.

In this paper we have considered the bulk form of the free energy and its
surface (order~$1$) correction.  In a future publication we will study the
finite-size (order~$1/N$) corrections at criticality, and hence derive the
central charges and conformal weights of the ABF models with fixed boundaries.

\bigskip
After this work was completed, we received the preprints \cite{BZ95} and
\cite{ZB95}, in which the authors derive surface critical exponents of the
eight-vertex and ABF models using methods similar to ours.

\section*{Acknowledgements}

We thank Ole Warnaar for helpful discussions.  This research is
supported by the Australian Research Council.

\pagebreak

\end{document}